\documentclass[a4paper]{jpconf}
\usepackage{graphicx}
\usepackage{amsmath}
\usepackage{amssymb}
\newcommand{\MeV}{\text{MeV}}
\newcommand{\GeV}{\text{GeV}}
\newcommand{\fm}{\text{fm}}
\newcommand{\ee}{\text{e}}

\begin{document}
\title{Quarkonia production in a Langevin approach}

\author{C Greiner, N Krenz and H van Hees}

\address{Institut f{\"u}r theoretische Physik, Goethe-Universit{\"a}t
  Frankfurt am Main, Max-von-Laue-Stra{\ss}e 1, 60438 Frankfurt,
  Germany}

\ead{krenz@th.physik.uni-frankfurt.de}

\begin{abstract}
  We aim to describe the process of dissociation and recombination of
  quarkonia in the quark-gluon plasma. Therefore we developed a model
  which allows to observe the time evolution of a system with various
  numbers of charm-anticharm-quark pairs at different temperatures. The
  motion of the heavy quarks is realized within a Langevin approach. We
  use a simplified version of a formalism developed by Blaizot et al. in
  which an Abelian plasma is considered where the heavy quarks interact
  over a Coulomb like potential. We have demonstrated, that the system
  reaches the expected thermal distribution in the equilibrium limit.
\end{abstract}

\section{Introduction}

Heavy quarks are an important tool for the investigation of the
quark-gluon plasma (QGP). They are primarily produced in the primordial
hard collisions of the heavy ion collision, and their number is
conserved until the hadronic freezout. Therefore heavy quarks carry
information about the whole evolution of the QGP. Especially the
surviving probability of heavy-quark bound states such as
$\text{J}/\psi$ or $\Upsilon$ can give an insight to the medium
properties.

The potential between two heavy quarks is screened by the surrounding
medium. As suggested long ago, the suppression of $\text{J}/\psi$ could
be an evidence for the formation of the deconfined state
\cite{MatsuiSatz}. Higher temperatures should lead to larger screening
effects with a full suppression of $\text{J}/\psi$ at very high beam
energies. The predicted suppression was found at the SPS at CERN
\cite{Baglin} but measurements at RHIC at higher beam energies did not
show an increase of the suppression \cite{Abelev}. To explain the
results the process of recombination of $\text{J}/\psi$ inside the
medium has been suggested. At higher beam energies the number
of initially produced heavy quarks is larger, leading to a higher
possibility that a heavy quark which propagates through the medium meets
a partner to form a new quarkonium state. The theoretical investigation
of recombination processes is therefore necessary to predict the number
of $\text{J}/\psi$-mesons measured in the experiments.

The comparatively large masses of heavy quarks makes their motion
accessible by Langevin dynamics \cite{Svetitsky}. The forces that act on
the charm quarks by using the Langevin equation are a drag force and
random momentum kicks due to collisions with the medium particles. To
enable the formation of bound states we add a potential between the
heavy quarks, that leads to a attractive force between charm and
anti-charm quarks.

In section \ref{formalism} we will explain the formalism and the
parameters that we have used in our simulation. The current results are
presented in section \ref{results} while section \ref{outlook} gives an
outlook on further applications for a description of heavy quarkonia in
heavy-ion collisions.

\section{Formalism}
\label{formalism}
\begin{figure}[h]
\begin{minipage}{0.49\textwidth}
\includegraphics[width=\linewidth]{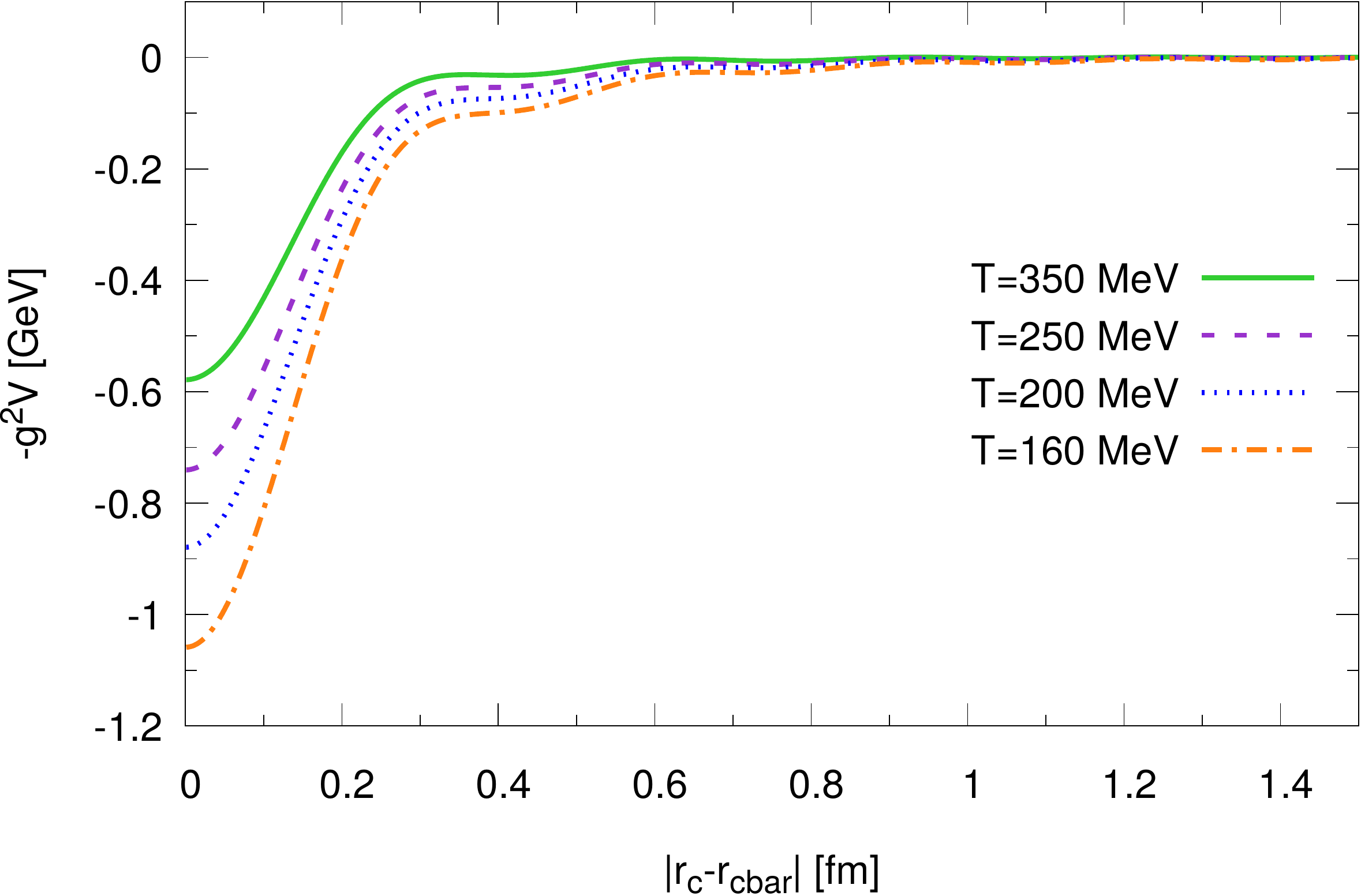}
\caption{\label{potential}The charm-anti-charm-quark pair potential for
  different temperatures.}
\end{minipage}\hfill
\begin{minipage}{0.48\textwidth}
\includegraphics[width=\linewidth]{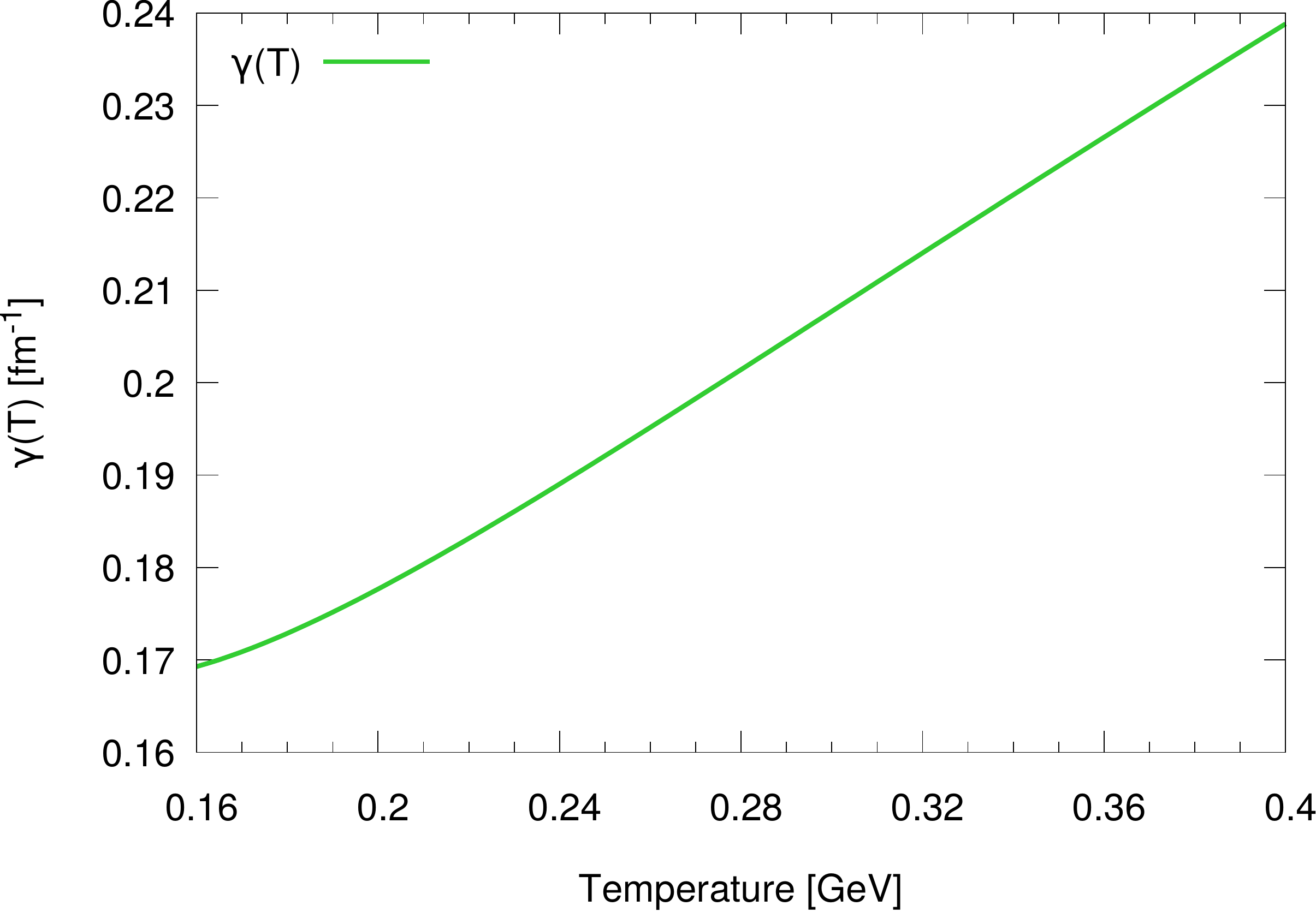}
\caption{\label{gamma-fig}The friction coefficient for different temperatures.}
\end{minipage} 
\end{figure}

For the realization of the heavy-quark motion we adopt the formalism by
Blaizot et al. \cite{Blaizot}. In this description the heavy-quark
interaction is reduced to an Abelian model, which means that confinement
and color effects are neglected. The Langevin update rules for the
coordinate $\boldsymbol{x}$ and the momentum $\boldsymbol{p}$ of a heavy quark
with mass $M$ read
\begin{alignat}{2}
&
\frac{\mathrm{d}\boldsymbol{r}}{\mathrm{d}t}=\frac{1}{2M}\boldsymbol{p}, \\
&\frac{\mathrm{d}\boldsymbol{p}}{\mathrm{d}t}=-\gamma
\boldsymbol{p}+F\left(\boldsymbol{r}-\boldsymbol{\bar{r}}
\right)+\sqrt{2MT\gamma \Delta t}\boldsymbol{\rho}, 
\end{alignat}
where $\gamma$ is the friction coefficient due to the interaction with
the medium, $F\left(\boldsymbol{r}-\boldsymbol{\bar{r}} \right)$ is the
force resulting from the heavy-quark potential, $T$ is the temperature
of the medium, and $\boldsymbol{\rho}$ are Gaussian normal-distributed
random numbers. For the pertinent diffusion coefficient the usual
Einstein dissipation-fluctuation relation has been employed. The
quark-anti-quark potential is given by a screened Coulomb potential with
a cut-off for large momenta at small distances. Following \cite{Blaizot}
the cut-off is taken to be $\Lambda=4$ GeV. The potential for different
temperatures is displayed in Figure \ref{potential}. The drag force in
\cite{Blaizot} contains a dependence on the distance between the heavy
quarks. For simplicity we neglect this dependence in our simulation and
use a constant drag value. In a numerical calculation a cut-off is also
necessary for the friction. With the same cut-off as for the potential
the drag-coefficient is given by
\begin{equation}
\label{gamma2}
\gamma = \frac{m_{\text{D}}^2 g^2}{24\pi M}\left[
  \mathrm{ln}\left(1+\frac{\Lambda ^2}{m_{\text{D}}^2}\right)
  -\frac{\frac{\Lambda ^2}{m_{\text{D}}^2}}{\frac{\Lambda ^2}{m_{\text{D}}^2}+1}\right], 
\end{equation}
where $m_{\text{D}}$ is the Debye screening mass, defined as
$m_{\text{D}}^2 = \frac{4}{3}g^2T^2$, which is the perturbative
expression for a two-flavor quark-gluon plasma. The gauge coupling $g$
is given by the relation \cite{Let}
\begin{equation}
g^2=4\pi \alpha_s=\frac{4\pi \alpha_s(T_C)}{1+C\mathrm{ln}\left( \frac{T}{T_C}\right) },
\end{equation}
with
\begin{equation*}
C=0.76, \quad T_C=160 \, \mathrm{MeV}, \quad \alpha_s(T_C)=0.5,
\end{equation*}
where $T_C$ is the critical temperature.  The value of $\gamma$ at
different temperatures can be seen in Figure \ref{gamma-fig}. The
charm-quark mass is set to $M=1.8 \, \GeV$, which within this model
results in binding energies of $\bar{\text{c}} c$ pairs leading to
bound-state masses close to typical charmonium masses.

\section{Results}
\label{results}

\begin{figure}[h]
\begin{minipage}[t]{0.49 \textwidth}
\includegraphics[width=\linewidth]{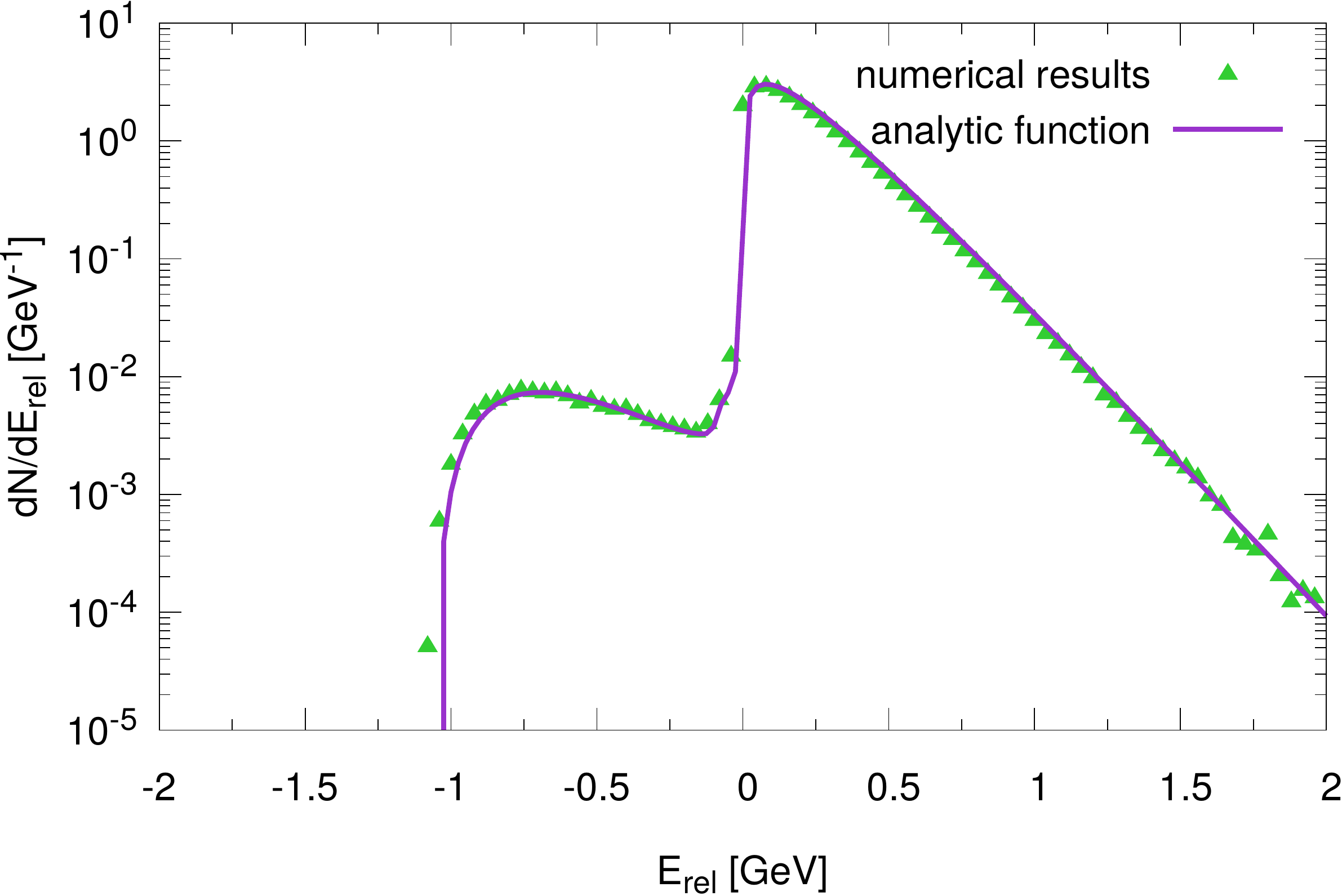}
\caption{\label{energydist}Pair-energy distribution at equilibrium. The
  simulation is done assuming a cubic box with boxsize $8 \, \fm$ at a
  temperature of  $T=160 \, \MeV$.}
\end{minipage} \hfill
\begin{minipage}[t]{0.49 \textwidth}
\includegraphics[width=\linewidth]{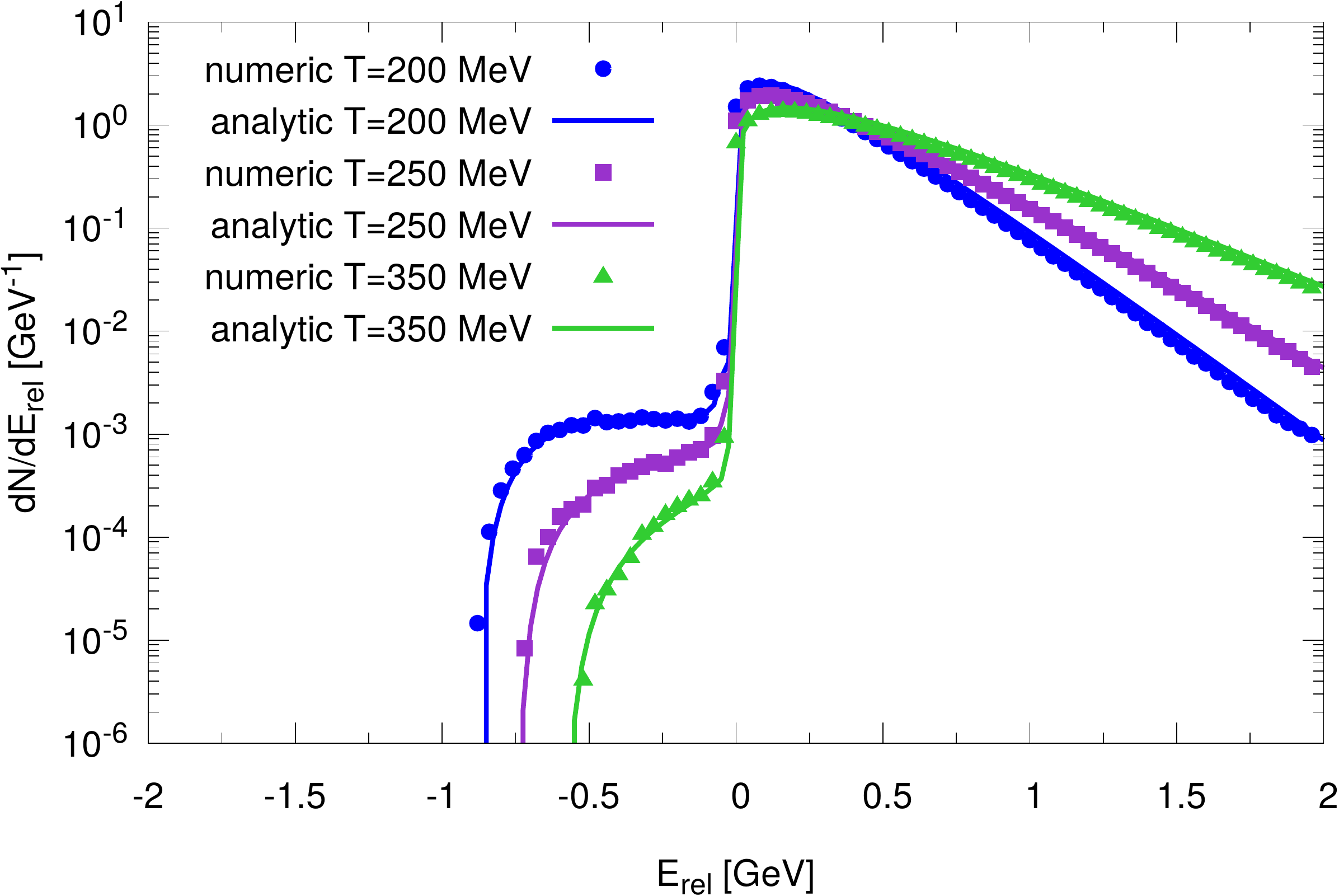}
\caption{\label{Tempanalytic} Change of the equilibrium distribution for
  different temperatures. Due to stronger screening effects higher
  temperatures lead to a smaller yield of bound states.}
\end{minipage} 
\end{figure}

First the simulation was tested with a single charm-anticharm pair in
the medium. The following simulations are done inside a cubic box with
side-length 8 fm. At the box boundary the particles are reflected. In
case a bound state hits the boundary, we reflect the center-of-mass
because we do not want to destroy the binding. To define bound states we
use the classical condition that two objects are bound if the energy of
the pair is negative. To check this we calculate the relative energy
$E_{\mathrm{rel}}$ of the pair, which means subtracting the
center-of-mass energy from the total energy. After the system reaches
equilibrium the distribution of the relative energy should be given by
the classical density of states
\begin{equation}
\frac{\mathrm{d}N}{\mathrm{d}E_{\mathrm{rel}}}=C\int_{\mathbb{R} ^3}
\mathrm{d}^3\boldsymbol{r} \int_{\mathbb{R} ^3} \mathrm{d}^3
\boldsymbol{p}_{\mathrm{rel}} \delta
(E_{\mathrm{rel}}-H_{\mathrm{rel}})\mathrm{exp}\left(
  -\frac{H_{\mathrm{rel}}}{T} \right),
\end{equation}
where $H_{\mathrm{rel}}$ is the Hamiltonian of the pair and $C$ is a
normalization constant.  We have solved these integrals numerically for
a sphere of radius $R$ with the same volume as our box which leads to
\begin{equation}
\label{dNdEanalytic} 
\frac{\mathrm{d}N}{\mathrm{d}E_{\mathrm{rel}}}=(4\pi)^2 (2\mu)^{3/2} C
\int_0^{R} \mathrm{d}r r^2 \sqrt{E_{\mathrm{rel}}-V(r)}
\mathrm{exp}\left(-\frac{E_{\mathrm{rel}}}{T}\right),
\end{equation}
and compare the results with those obtained in the simulation. The
results are shown in Figure \ref{energydist}. For this plot both curves
are normalized to one. We see that the numerical calculation perfectly
fits to the analytic function.
We also found this agreement for different temperatures, as can be seen
in Figure \ref{Tempanalytic}. For higher temperatures the number of
bound states in equilibrium decreases due to stronger screening
effects. The different shapes of the box volume, which is a cube in the
numerical simulation and a sphere in the analytic calculation seems not
to change the result. This indicates that we choose the volume large
enough so that the boundary effects can be neglected.

\begin{figure}[h!]
\begin{minipage}[t]{0.49\textwidth}
\includegraphics[width=\linewidth]{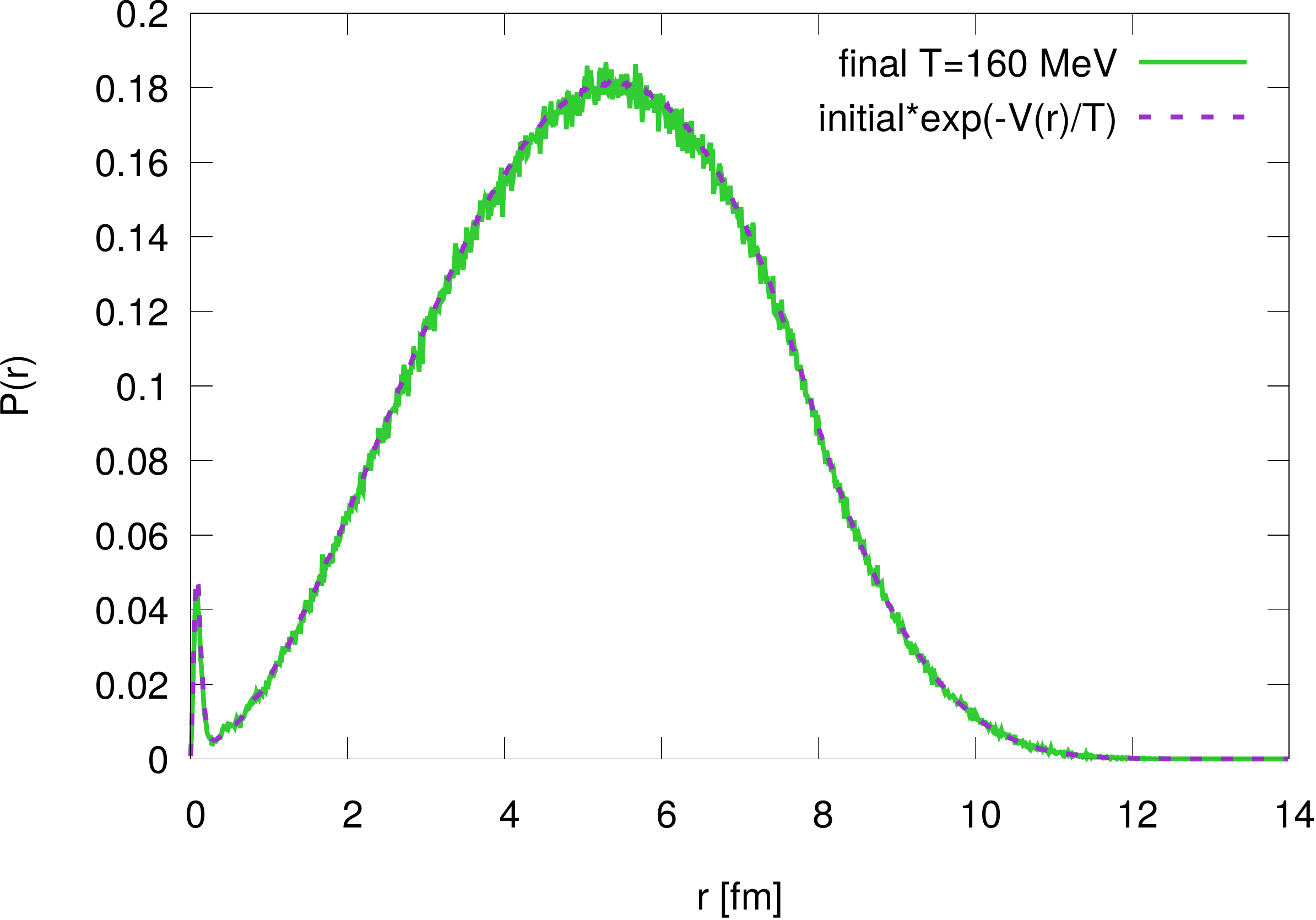}

\end{minipage}\hfill
\begin{minipage}[t]{0.49 \textwidth}
  \includegraphics[width=\linewidth]{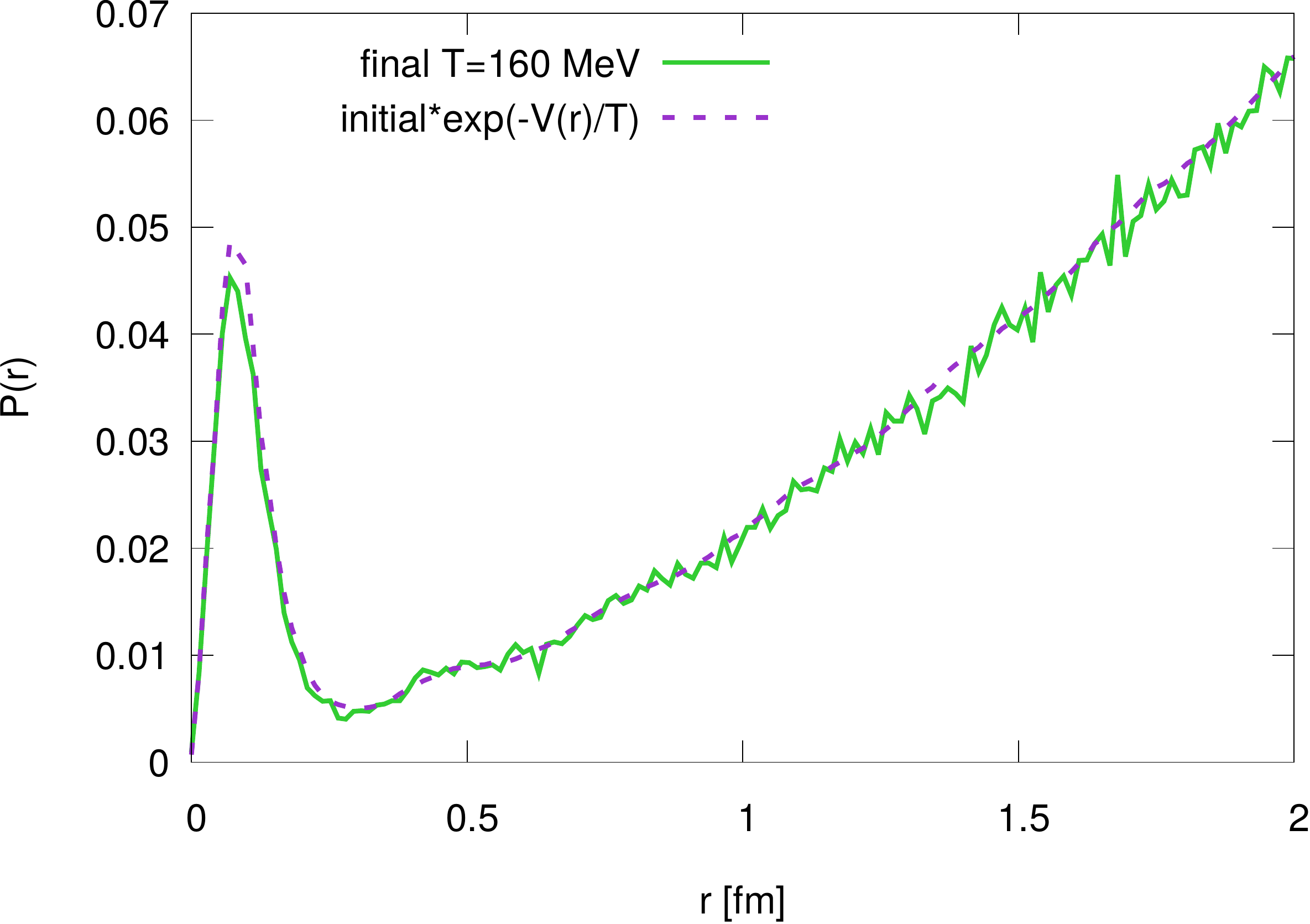}
\end{minipage}
  \caption{\label{Rhofull}Probability distribution for a single charm
  anti-charm pair to have the distance $r$ when the system is
  in equilibrium. The distribution is calculated in the numerical
  simulation for a single pair in a cubic box with side length 8 fm at
  temperature $T=160$ MeV (green solid line) compared to the analytic
  expectation cf.\ Eq.\ (\ref{rhoanalytic}) (violet dashed line). The
  right panel zooms into the small-distance region to illustrate the
  effect of the formation of bound states, leading to a pronounced peak.}
\end{figure}

To illustrate the attraction between the charm and the anti-charm quark
we plot the distance distribution of a single pair at temperature
$T=160 \, \MeV$. The results are shown in Figure \ref{Rhofull}, where
the green (solid) line represents the numerical simulation. The data are
taken in the long-time limit when the system is
equilibrated. Analytically the distribution can be calculated by
evaluating
\begin{equation}
  \label{rhoanalytic} 
P(r)=\frac{1}{R^6}\int_{R^3} \mathrm{d}^3\boldsymbol{r}_1 \int_{R^3}
\mathrm{d}^3\boldsymbol{r}_2 \; \delta\left(
  r-|\boldsymbol{r}_1-\boldsymbol{r}_2|\right)
\ee^{-\frac{V(r)}{T}}, 
\end{equation}
where $R^3$ is the volume of the box and $V(r)$ is the potential between
the heavy quarks.
The violet (dashed) line represents a Monte-Carlo evaluation of (\ref{rhoanalytic}).
The numerical and the analytic function are in good agreement.

We have also investigated the time evolution of the bound states at a
temperature of $T=160 \, \MeV$ for
two initial conditions.

\begin{figure}[h!]
\begin{minipage}[t]{0.49\textwidth}
  \includegraphics[width=\linewidth]{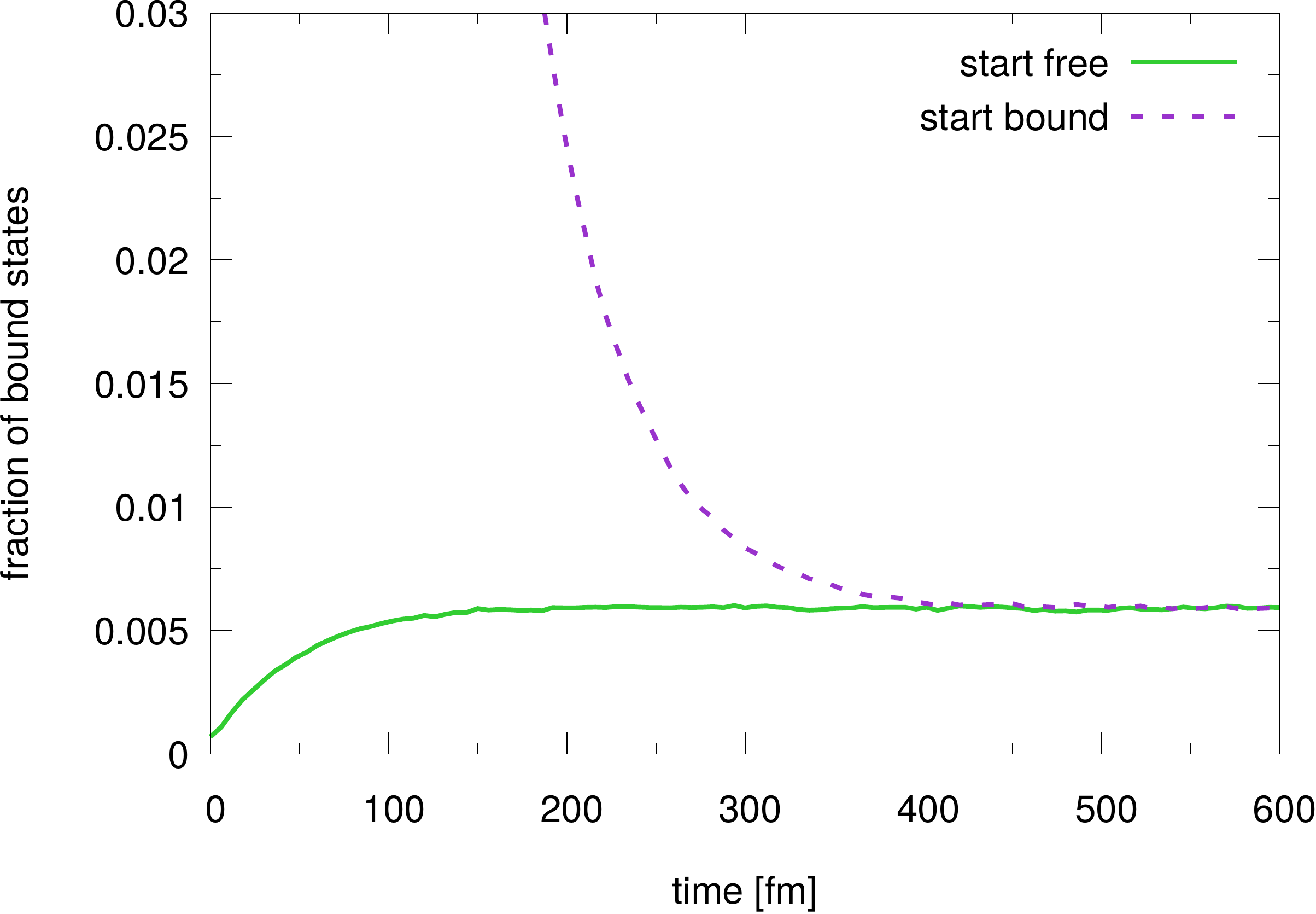}\hfill
  \caption{\label{freebound}Time evolution of the fraction of bound
    states with the charm-anticharm pairs produced initially bound
    (violet dashed line) or placed randomly inside the box (green solid
    line). Both curves lead to the same equilibrium limit but the
    equilibration time is longer in case of initial bound states.}
\end{minipage}\hfill
\begin{minipage}[t]{0.49\textwidth}
  \includegraphics[width=\linewidth]{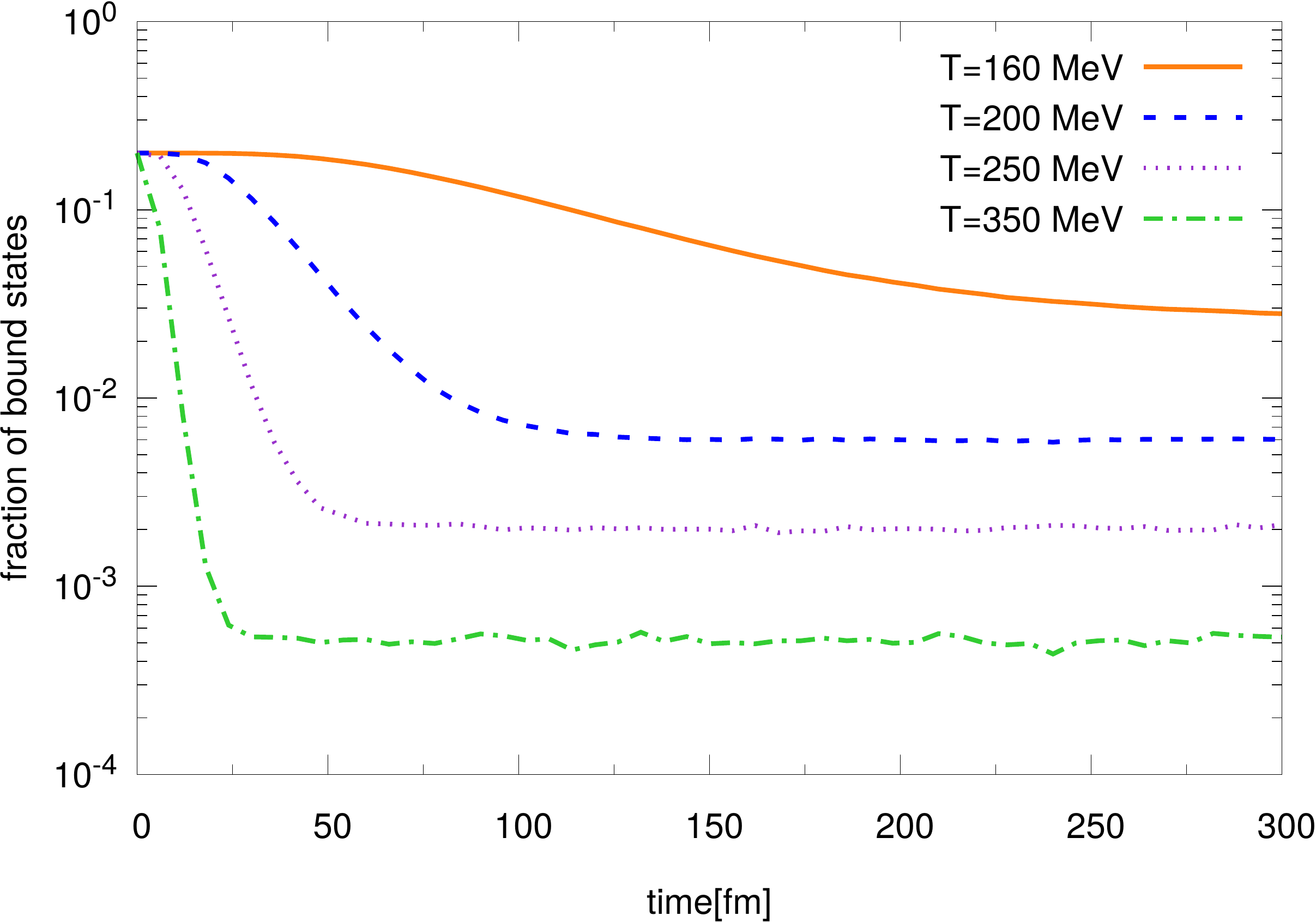}\hspace{2pc}%
  \caption{\label{TimeevoTemp}Time evolution of
    the fraction of bound states for different temperatures. Higher
    temperatures lead to a less amount of bound states and to a faster
    equilibration.}
\end{minipage} 
\end{figure}
In the first simulation the charm and anticharm quarks are initially
randomly placed inside the box to obtain the time evolution for the
formation of bound states. For the second simulation the heavy quarks
are initially created as bound states with a pair energy of
$-700 \, \MeV$, which approximately corresponds to the maximum of the
peak on the left panel of Figure \ref{energydist}. The momenta are set
back-to-back with a relative velocity that is taken from a Maxwell
distribution with its center at the average value of typical relative
velocities of charmonium $v_0^2\sim 0.3$ \cite{Bodwin}. As shown in
Fig.\ \ref{freebound}, in both cases the fraction of bound states in the
system equilibrates to the same value as expected. At this point the
dissociation and recombination rates are equal. The time till the
equilibrium value is reached is much larger if the pairs are initially
created as bound states. This is due to the strong binding between the
quarks. Separating an existing pair requires a large energy transfer
from the medium. We notice, that in both cases we found very long
timescales for the equilibration.

To see the influence of the medium's temperature on the equilibration
time we calculated the time evolution at different temperatures. In this
simulation we produced five charm-anticharm pairs, initially created as
bound states. The results are displayed in Figure \ref{TimeevoTemp}.

Larger screening effects at higher temperatures lead to a smaller
fraction of bound states at equilibrium. Also the equilibration time
decreases. This is expected, because the friction coefficient increases
with temperature. The momentum and therefore also the energy transfer in
the collisions with the light medium constituents is higher, leading to
a faster dissociation of a bound state.

\begin{figure}[h!]
  \begin{minipage}[b]{0.49\textwidth}
    \includegraphics[width=\linewidth]{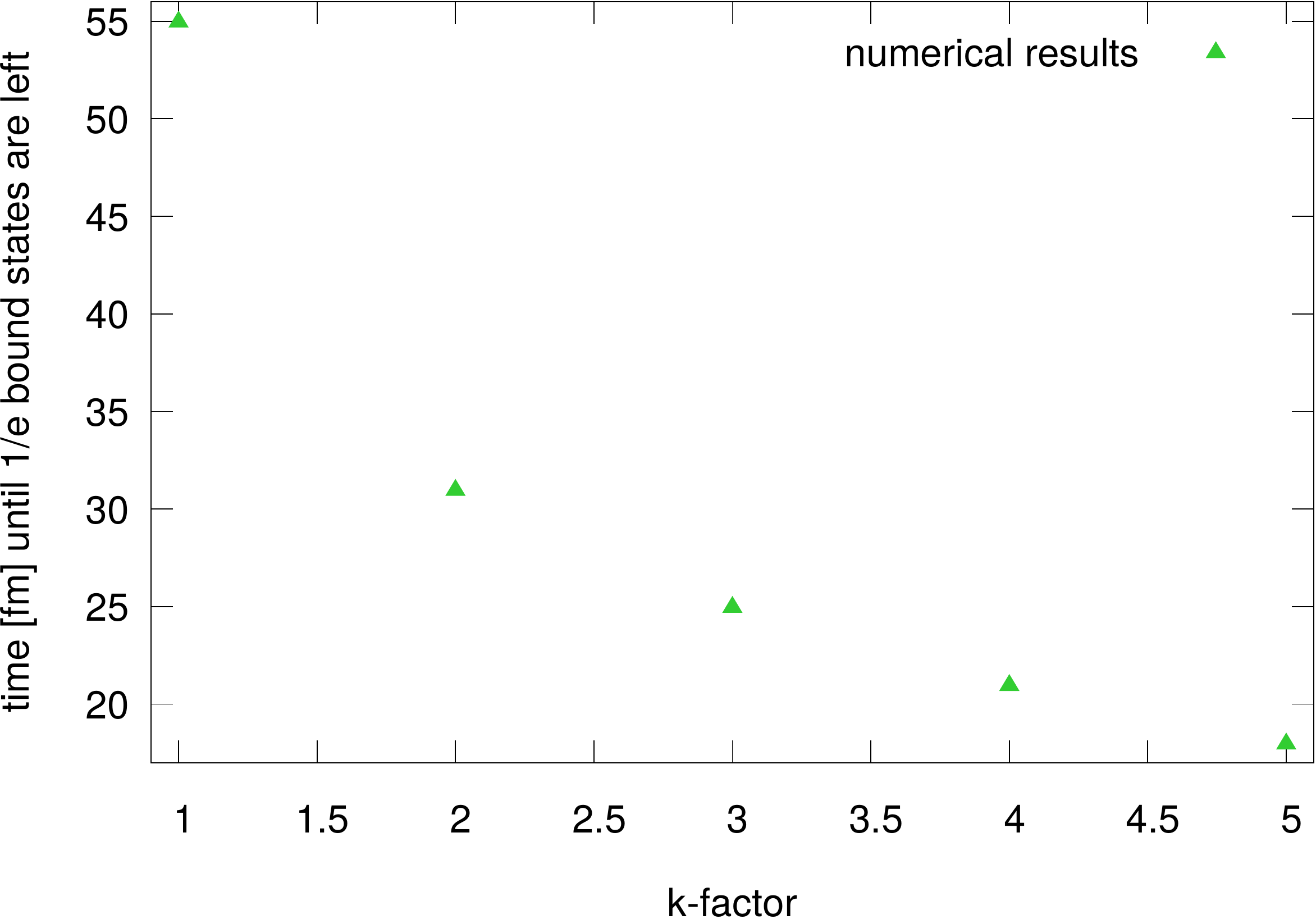}\hfill
  \end{minipage} \hfill
  \begin{minipage}[b]{0.49\linewidth}
    \caption{\label{1etel} Equilibration time as a function of different
      multiples $k$ of the friction coefficient $\gamma$, defined as the
      time after which the fraction of bound states has decreased to
      $1/\text{e}$ of its initial value.}
\end{minipage}
\end{figure}

To investigate the relation between the relaxation time and the friction
coefficient we have run the simulation for different multiples
$k \gamma$. For this simulation we created a single pair as an initial
bound state and calculated the time until there is only a fraction of
$1/\mathrm{e}$ of the initial pairs left. The results are shown in
Figure \ref{1etel}.

According to the kinetic equilibration time $1/\gamma$ we expect that
the decrease of the time also depends on $1/\gamma$. In our simulation
the decrease is a bit slower but has the expected shape. To explain the
large time scales we need to include chemical effects on the
equilibration time in addition to the kinetic scales. Still the
behavior of the relaxation time for different friction coefficients is
approximately consistent with the underlying theory.

To check if our simulation is also in accordance with an equilibrated
thermodynamic system, we calculated the particle number using the grand
canonical partition function. In a classical approximation the particle
number is given by
\begin{equation}
  \label{numberJpsi}
  N=gV\left( \frac{mT}{2\pi} \right) ^{\frac{3}{2}} \ee^{-m/T}\ee^{\mu /T},
\end{equation}
where $g$ is the degeneracy factor of the particles, $V$ is the volume
of the system, $T$ is the temperature, $m$ is the mass of the particle,
and $\mu$ the chemical potential. The factor $\ee^{\mu /T}$ is also
called the fugacity and will in the following be denoted with
$\lambda$. Since the dissociation and recombination of a $J/\psi$-meson
is a process of the kind
\begin{equation*}
J/\psi 	\rightleftarrows c + \bar{c},
\end{equation*}
the chemical potentials are connected by $\mu_{J/\psi}=2\mu_c$ which
means $\lambda_{J/\psi}=\lambda_c^2$. The number of $J/\psi$ in the
system can then be calculated by
\begin{equation}
  \begin{split}
    \label{numberJpsi2}
    N_{J/\psi}&=\lambda_c^2g_{J/\psi}V\left(\frac{m_{J/\psi}T}{2\pi}\right)^{\frac{3}{2}}\ee^{-m_{J/\psi}/T}\\
    &=N_c^2\left(\frac{g_{J/\psi}}{g_c^2}\right) \frac{1}{V}\left(
      \frac{m_{J/\psi}}{m_c^2} \right)^{\frac{3}{2}}\left(
      \frac{2\pi}{T} \right)^{\frac{3}{2}}\ee^{(2m_c-m_{J/\psi})/T}
\end{split}
\end{equation}
Divided by the initial number of $c\bar{c}$ pairs equation
(\ref{numberJpsi2}) gives the fraction of bound states in the
equilibrated system. We compared the results from our simulation with
those obtained by using equation (\ref{numberJpsi2}) for three different
box sizes in Table \ref{tab}. We find that both values are in the same
order of magnitude.
\begin{table}[h]
\caption{\label{tab}Comparison between the fraction of bound states
  obtained with our simulation and that calculated using equation
  (\ref{numberJpsi2}).} 
\begin{center}
\begin{tabular}{lll}
\br
Volume&grand canonical fraction of $J/\psi$&numerical fraction of $J/\psi$\\
\mr
$8^3$ fm$^3$ &0.0066&0.0059\\
$10^3$ fm$^3$ &0.0035&0.0029\\
$12^3$ fm$^3$ &0.002&0.0017\\ 
\br
\end{tabular}
\end{center}
\end{table}

\section{Conclusion and Outlook}
\label{outlook}

We have developed a model that allows to investigate the time evolution
of a system with multiple pairs of charm and anti-charm quarks using the
Langevin equation. We could show that our model passes all equilibrium
tests and that the bound-state properties in a box are consistent with
the expectations for a grand canonical ensemble.

Various extensions are possible to improve the model. First we want to
include a distance dependent friction coefficient as used in
\cite{Blaizot} to replace our constant $\gamma$ in equation
(\ref{gamma2}) that we chose for simplicity. To simulate a heavy-ion
collision with a hot and dense state at the beginning that expands and
cools down until the hadronic freezout we want to describe the medium
evolution as an expanding fireball. This enables us to observe the
number of $J/\psi$'s at the different stages of a collision. We also
want to explore different quarkonia potentials. Especially potentials
that include effects from strong interaction such as confinement are
important to obtain a more realistic description. The influence of the
color charges carried by the charm and anticharm quarks have to be
considered too. Also the classical picture should be replaced by quantum
mechanical calculations. A first step could be to decide whether a bound
state is created by using the Wigner function as in \cite{Shuryak}. In
addition we have to replace the continuous energy distribution of the
charm-anti-charm pairs with a quantized one that allows only certain
energy levels. The energy levels could be comparable with the different
excitation states of charmonium to calculate the respective
abundance. The long-time goal of this project is to obtain a full
in-medium quantum Langevin treatment of quarkonia.

\end{document}